\documentclass[aps,prl,twocolumn,groupedaddress]{revtex4-2}

\usepackage{graphicx}
\usepackage{amsmath}

\usepackage{xcolor}
\usepackage{soul}
\usepackage{bm}
\usepackage{svg}

\newcommand{\ttau}{\bm{\tau}}

\begin{document}

\title{The Effect of Internal Damping on Locomotion in Frictional Environments}

\author{Brian Van Stratum}
\author{Jonathan Clark}
\author{Kourosh Shoele}
 \email{kshoele@eng.famu.fsu.edu}

\affiliation{Department of Mechanical Engineering, FAMU-FSU College of Engineering, Tallahassee, FL, 32310, USA}

\date{\today}

\begin{abstract}
The gaits of undulating animals arise from a complex interaction of their central nervous system, muscle, connective tissue, bone, and environment. As a simplifying assumption, many previous studies have often assumed that sufficient internal force is available to produce observed kinematics, thus not focusing on quantifying the interconnection between muscle effort, body shape, and external reaction forces. This interplay, however, is critical to locomotion performance in crawling animals, especially when accompanied by body viscoelasticity. Moreover, in bio-inspired robotic applications, the body's internal damping is indeed a parameter that the designer can tune. Still, the effect of internal damping is not well understood. This study explores how internal damping affects the locomotion performance of a crawler with a continuous, visco-elastic, nonlinear beam model. Crawler muscle actuation is modeled as a traveling wave of bending moment propagating posteriorly along the body. Consistent with the friction properties of the scales of snakes and limbless lizards, environmental forces are modeled using anisotropic Coulomb friction. It is found that by varying the crawler body's internal damping, the crawler's performance can be altered, and distinct gaits could be achieved, including changing the net locomotion direction from forward to back. We will discuss this forward and backward control and identify the optimal internal damping for peak crawling speed.
\end{abstract}

\maketitle

\section{Introduction}

The locomotion of long slender crawlers arises from the interaction of muscles and ligaments and external forces such as Coulomb friction and viscous drag from fluids. The mechanism of movement for long thin organisms has been the subject of many previous investigations \cite[e.g.][]{gray1946mechanism,jayne1986kinematics,gasc1989propulsive}. Long slender organisms across length scales and various media have adopted a variety of locomotion gaits. Terrestrial snake locomotion, for example, can be classified into one of four main locomotion modes: lateral undulation, concertina, sidewinding, and rectilinear \cite{gray1946mechanism}. It is known that snakes switch between these modes subject to their environment as snakes are  motile in diverse media such as granular, aquatic, and aerial domains \cite{socha2002gliding,marvi2014sidewinding,jayne1988muscular}. From a robotics perspective, the hyper-redundant body morphology of snakes, which provides mastery of such varied terrain and stability, makes them excellent subjects for biologically inspired designs of snake robots, surgical devices, and targeted drug delivery mechanisms.  

\begin{figure}
\centering
\includegraphics[width=1\columnwidth]{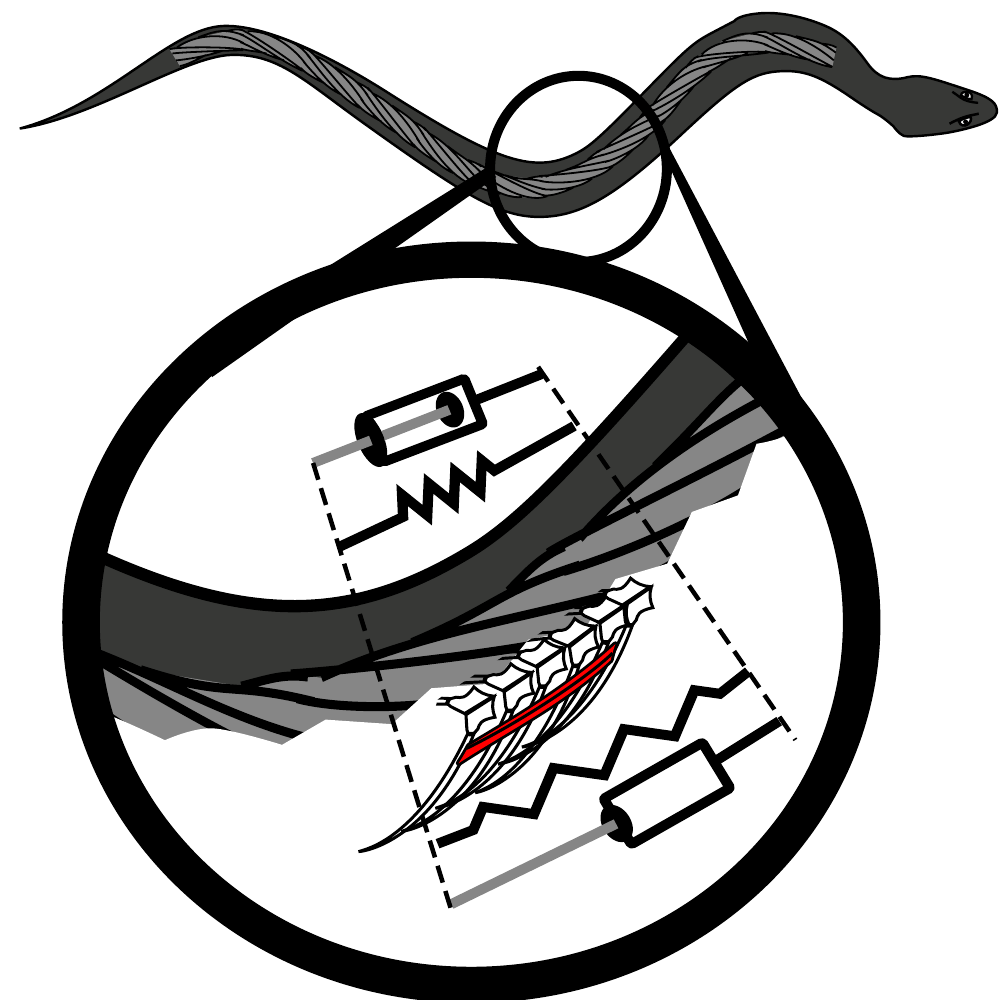}
\caption{Modeling a Snake-like organism as a visco-elastic beam}
\end{figure}

Many animals use lateral undulation to move in all sorts of environments. Examples include crawling snakes in grass, eels in water and sandfish in sedimentary environments such as sand \cite{maladen2009undulatory}. Further, some animals combine lateral undulation with other motility techniques, such as fracture and crack propagation\cite{dorgan2005burrow}, burrowing by sediment removal, and forms of two anchor crawling utilizing peristaltic waves\cite{dorgan2015biomechanics}. In addition to the utility of lateral undulation in various environments, it also occurs across a wide range of length scales, from reticulated pythons  all the way down to nematode worms, such as beet eelworms, which propagate waves of curvature from head to tail down their body to move in sand and soft mud\cite{wallace1958movement}. Since lateral undulation relies on a sliding interaction with the ground, friction is both essential and destructive to the crawler. Without friction, there can be no net forces driving locomotion, yet the friction results in energy dissipation at the crawler's skin and affects their cost of transport. Yet,  the transport cost of snake-like animals moving by sliding across the ground is comparable to birds and limbed creatures. Because of the applicability of lateral undulation in various  environments, across length scales and for various body morphologies, it forms the focus of this study. \citep{walton1990energetic,secor1992locomotor}.

Lateral undulation has also been employed in a new class of snake robots that are soft and compliant in nature\cite{shepherd2011multigait,onal2013autonomous,branyan2017soft}. These soft robots hold out the promise of highly customized robots whose body motion and dynamics can be programmed into the material of the body. Recent soft robotic snakes can mimic their animal prototypes and use curvature waves for locomotion \cite{qi2020novel}. These robots can be used for complex search operations, inspection, and maintenance in congested and chaotic environments.

\begin{figure*}
\centering
\includegraphics[width=2\columnwidth]{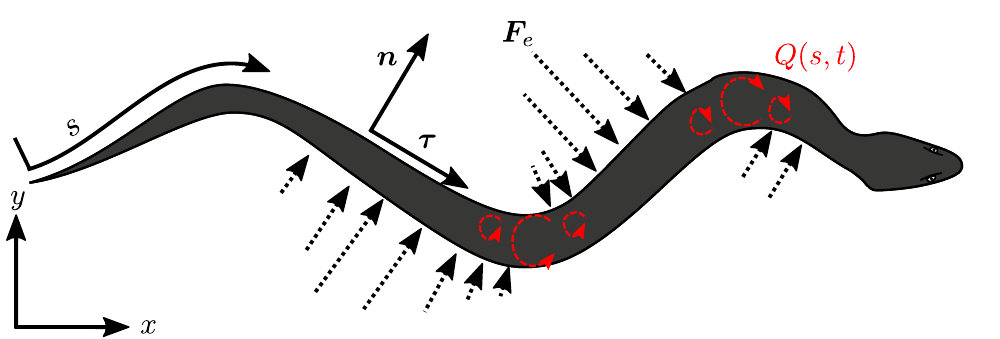}
\caption{Free body diagram of the local coordinate system, environmental force density, and internal torque density.}
\label{Fig: mathDiagram}
\end{figure*}

Different techniques have been proposed to model the kinematics and dynamics of crawling robots \citep{transeth2009survey}. These models represent the interface between the crawler and its environment as either no-slip or frictional slip conditions. For the no-slip condition, the ground interaction is modeled as a non-holonomic constraint at every point on the body's surface. This constraint treats a snake belly as a flexible ice skate. In the slipping condition, on the other hand, the environmental forces are represented with specific frictional or resistive models, and the body is allowed to move laterally relative to the environment. Such a model has been frequently used for studying the slithering snake motion, notably \citep{hu2009mechanics,guo2008limbless}. Slithering snakes distribute their weight away from the points of maximum curvature along their bodies in a locomotion strategy known as sinus lifting. Hu et al. \cite{hu2009mechanics} advanced the understanding of slithering locomotion by introducing a model explaining the critical role of weight distribution in snake locomotion. This sinus-lifting action has also been leveraged to improve the functionality of snake robots without wheels \cite{liljeback2012review,alben_efficient_2022}.   

Researchers have used frictional models to study the optimal locomotion problem, studying the interaction of a crawler's body with a frictional environment. \cite{hu2009mechanics,alben2013optimizing,alben2019efficient}. Alben studied a planar crawler subject to sliding friction in the forward, transverse, and backward directions and used a quasi-Newton method to find the time and space curvature trajectory that optimizes locomotion efficiency \cite{alben2013optimizing}. Among other things, Alben discovered that the optimal direction of locomotion changes as a function of environmental friction. Later, it was shown that the optimal gait is mainly a function of the transverse and forward friction coefficient ratio, and optimal crawling performance is in the direction of the curvature wave propagation when transverse friction is low and opposite that when the transverse friction is large \cite{alben2019efficient}. Alben referred to these as ``direct" and ``retrograde" curvature waves. Guo et al. \cite{guo2008limbless}  investigated the coupled effect of internal and external parameters, including the viscous and elastic moment of a visco-elastic crawler with the non-holonomic, no-slip velocity constraint. They found that parameters in the external environment mainly determine a crawler's body shape, and the speed of the crawler is primarily determined by internal parameters (exogenous and endogenous parameters, respectively.) McMillen found that damping plays a critical role in the swimming of Lamprey \cite{mcmillen2008nonlinear}. McMillen showed that muscle contraction in slender swimmers happens slightly after muscle activation, and this delay increases progressively from head to tail. McMillen related this observation to  the viscoelastic property of the tissue, the geometry of the swimmer, and the muscle dynamics, more so than the effect of fluid forces. This work indicates that damping is important in lateral undulation in general and significant for designing embodied intelligence into robotic systems. 

The role of body viscoelastic damping is especially important for pneumatically actuated soft robots. Damping is inherent to these robots, which are composed of materials with very low Young's moduli and include fluidic channels for actuation \cite{mosadegh2014pneumatic,chen2020soft}. By tuning the damping behavior of soft robot actuators, it is also possible to improve their performances \cite{di2019dynamic}. However, most of the prior research focused on the importance of internal compliance within the body of a crawler \cite{rollinson2013gait,whitman2016shape}. It has been demonstrated experimentally that by leveraging  the robot's flexibility, it is possible to increase the snake robot's ability to traverse two and three-dimensional obstacles\cite{fu2020robotic,schiebel2020robophysical}. However, little is known about how internal damping affects the locomotion of long slender organisms and robots relative to their stiffness. The notable exception is the work done by Guo, which investigated the role of internal viscous damping on a crawler with the non-holonomic constraint\cite{guo2008limbless}. However, this non-holonomic constraint does not apply to the general locomotion case and particularly does not apply to soft bio-inspired robots where substantial lateral slipping motion is seen \cite{branyan2020snake}. 
In this study, we explore the effects of internal damping and its connection with the lateral friction coefficient on the crawling performance of a modeled slender system that can serve as a guide for making decisions about the effect of damping in soft robot designs and also provide insight into bio locomotion.

\section{Model}
We study the locomotion of long slender organisms using a three-dimensional nonlinear Euler–Bernoulli beam formulation originally proposed  by Tjavaras et al.\cite{tjavaras1996dynamics}.  This model has been employed before to study the dynamics of animal locomotion from swimming fish to climbing lamprey \cite{cheng1998continuous,zhu2011numerical,zhu2008propulsion}.  In the model,  an Euler-Lagrangian dual coordinate
system is employed to formulate the motion, namely the global coordinate system($x$,$y$,$z$)
and a local reference system with unit vectors $\ttau$,
 \textbf{n}, and \textbf{b} in the tangential, normal, and
bi-normal directions of the centerline of long cylindrical locomotor, respectively. The two coordinate
systems are related by
\begin{equation}
\left[ \begin{gathered}
\ttau  \hfill \\
\bm{n} \hfill \\
\bm{b} \hfill \\
\end{gathered}  \right] = \mathbf{C} \cdot \left[ \begin{gathered}
\bm{i} \hfill \\
\bm{j} \hfill \\
\bm{k} \hfill \\
\end{gathered}  \right], \label{a1-eq1}    
\end{equation}
where $\bm{i}$, $\bm{j}$, and $\bm{k}$ are unit vectors in
$x$, $y$, and $z$ directions, respectively. \textbf{C} is
an orthogonal rotation tensor representing the change of orientation from the
global to the local reference frames. \textbf{C} is constructed by
using the singularity-free method of Euler parameters ${\beta _0}$,
${\beta _1}$, ${\beta_2}$, and ${\beta _3}$ to describe  the rotation based on a single angle $\vartheta$ and a principle direction $\bm{\ell}$ as:
{\small
\begin{equation}
\label{a1-eq3}  
C=\left[ \begin{gathered}
\beta_0^2 + \beta_1^2 - \beta_2^2 - \beta _3^2\quad \;2\left(
{{\beta _1}{\beta _2} + {\beta_0}{\beta_3}} \right)\quad 
2\left( {{\beta _1}{\beta _3} - {\beta _0}{\beta _2}} \right) \hfill \\
\;2\left( {{\beta _1}{\beta _2} - {\beta _0}{\beta _3}} \right)\quad
\beta_0^2 - \beta _1^2 + \beta _2^2 - \beta _3^2\quad 2\left(
{{\beta_2}{\beta _3} + {\beta _0}{\beta _1}} \right) \hfill \\
\;2\left( {{\beta _1}{\beta _3} + {\beta _0}{\beta _2}} \right)\quad
2\left( {{\beta _2}{\beta _3} - {\beta _0}{\beta _1}}
\right)\quad \beta _0^2 - \beta _1^2 - \beta _2^2 + \beta _3^2 \hfill
\\
\end{gathered}  \right]  
\end{equation}
}       
where
\begin{equation}
\left[ \begin{gathered}
{\beta _0} \hfill \\
{\beta _1} \hfill \\
{\beta _2} \hfill \\
{\beta _3} \hfill \\
\end{gathered}  \right] = \left[ \begin{gathered}
\cos \left( {\vartheta /2} \right) \hfill \\
{\ell _x}\sin \left( {\vartheta /2} \right) \hfill \\
{\ell _y}\sin \left( {\vartheta /2} \right) \hfill \\
{\ell _z}\sin \left( {\vartheta /2} \right) \hfill \\
\end{gathered}  \right], \label{a1-eq2}    
\end{equation}   
Dynamic equations are derived by imposing the
conservation of translational and angular momenta of an infinitesimal
segment of the body. We have
\begin{equation}
m\left( {\frac{{\partial {\bm{V}}}}{{\partial
t}} + {\bm{\omega }} \times {\bm{V}}} \right) = \frac{{\partial
{\bm{T}}}}{{\partial s}} + {\bm{\Omega }} \times {\bm{T}} +
\left( {1 + \varepsilon } \right) {{{\bm{F}}_e}
}  \label{a1-eq4}    
\end{equation}         
and
\begin{equation}
\frac{{\partial {\bm{M}}}}{{\partial s}} +
{\bm{\Omega }} \times {\bm{M}} + \left( {1 + {\varepsilon}}
\right)^3\ttau  \times {\bm{T}} +\left( {1 + {\varepsilon}}
\right)^2 \bm{Q} = 0, \label{a1-eq5}    
\end{equation}
where $m$ is the mass per unit length.
$s$ is the distance from this point to the tail
end along the unstretched crawler. ${\bm{V}}(s,t) =
{V_\tau }\ttau  + {V_n}{\bm{n}} + {V_b}{\bm{b}}$ and
${\bm{\omega }}(s,t) = {\omega_\tau }\ttau  +
{\omega _n}{\bm{n}} + {\omega _b}{\bm{b}}$ are the
translational and angular velocities, respectively.
${\bm{T}}(s,t) = {T_\tau }\ttau  +
{T_n}{\bm{n}} + {T_b}{\bm{b}}$ is the internal force
and ${\bm{M}}(s,t) = {M_\tau }\ttau  +
{M_n}{\bm{n}} + {M_b}{\bm{b}}$ is the internal moment.
$\varepsilon (s,t)$ is the axial strain and 
${\bm{\Omega }}(s,t) = {\Omega _\tau }\ttau  +
{\Omega _n}{\bm{n}} + {\Omega _b}{\bm{b}}$ is the
Darboux vector measuring the material torsion and curvatures of the
body centerline. ${{\bm{F}}_e}$ is the  frictional
force  density, and $\bm{Q}$ is the internal moment density generated by the body, both to be defined later.

One compatibility relation is required  to ensure a continuous body  configuration
in space and time,
\begin{equation}
\frac{{\partial \varepsilon }}{{\partial
t}}{{\ttau }} + \left( {1 + \varepsilon } \right){\bm{\omega
}} \times {{\ttau }} = \frac{{\partial {\bm{V}}}}{{\partial
s}} + {\bm{\Omega }} \times {\bm{V}}, \label{a1-eq6}    
\end{equation}                        

The internal forces and moments are related to the strain
$\varepsilon $ and the Darboux vector
${\bm{\Omega }}$ through the
constitutive relations ${T_\tau } = EA\varepsilon
$, ${M_\tau } = GJ{\Omega _\tau }$,
${M_n} = EI{\Omega _n}$,
${M_b} = EI{\Omega _b}$. By definition,
$A$ is the cross-sectional area. \textit{EI} and
\textit{GJ} are the bending and torsional stiffness, respectively. 
 Aggregate elastic Young’s modulus $E_{el}$ and visco-elastic
coefficient $D=\alpha E_{el}$ represents the combined contribution from all the passive elements during bending. This, in real animals, includes a viscoelastic  spine,  skin, white muscle, and the inactive part of red muscles. Mathematically,  linear viscoelasticity is introduced  by defining  $E=E_{el} (1+\alpha\, \partial/\partial t)$. The final system of equations is

\begin{equation}
\frac{{\partial {\bm{Y}}}}{{\partial s}} +
{\mathbf{H}}\frac{{\partial {\bm{Y}}}}{{\partial t}} + {\bm{P}}
= 0, \label{a1-eq7}    
\end{equation} 
where ${\bm{Y}} = {\left[ {\varepsilon
\;{T_n}\;{T_b}\;{V_\tau }\;{V_n}\;{V_b}\;{\beta _0}\;{\beta _1}\;{\beta
_2}\;{\beta _3}\;{\Omega _\tau }\;{\Omega _n}\;{\Omega _b}}
\right]^T}$. A detailed description of the matrices \textbf{H}
and $\bm{P}$ is provided by Tjavaras \textit{et al}. (1998).

Overall, thirteen boundary conditions are required at the two ends of
body. At the tail end, we apply seven boundary conditions: ${\beta _0} = 0$; $\bm{T}=0$ and $\bm{M}=0$.  Six
additional boundary conditions are applied at the head:
$\sum_i\beta_i^2  =
1$; $\bm{T}=0$; and ${M_n=0}$ and
${M_b=0}$. To solve Eq. (\ref{a1-eq7}), we
 divide the body into ${N_p} -
1$ segments, each with length $\Delta s$, by uniform 
distribution of points
${s_k}$ ($k = 1, \cdots
,{N_p}$) along the unstretched length of the body. An implicit box method is then applied to integrate the system of equations from the time step
${t_{i - 1}}$ to ${t_i} =
{t_{i - 1}} + \Delta t$ \cite{tjavaras1996dynamics}. The method is
second-order accurate in space and first-order accurate in time. A lower-order time integration scheme is used for  better numerical stability. Following the convergence study, $N_p=50$ is found to be sufficient for the current study. 

\subsection{Environmental Friction}
Snakes and other crawling animals possess body features such as scales that give their bodies higher coefficient of friction when sliding sideways relative to when they are sliding forward or backward \citep{hu2009mechanics,marvi2012friction,alben2013optimizing}. To replicate this property in the current mathematical model,  an anisotropic Coulomb friction model is used. In this model, the friction force is calculated in the body attached Lagrangian coordinate system as a function of the body velocity, $\bm{V}$,  as, 
\begin{align}
\bm{F}_{e} &= -
m\,g (\mu_{\tau}\,\hat{V}_{\tau}\ttau\,+\,\mu_n\,\hat{V}_{n}\bm{n})
\label{eq:EnvFriction}
\end{align}
where $\hat{V}_{\tau}$ and $\hat{V}_{n}$ are the regularized direction of velocity vectors along the body in the tangential and normal directions, respectively. They are defined as,
\begin{align}
(\hat{V}_{\tau},\hat{V}_{n}) &= \frac{
\left(V_{\tau},V_n\right)}{\sqrt{V^2_{\tau}+V^2_{n}+\epsilon^2}}
\label{eq:EnvFrictionb}
\end{align}
where $\epsilon$ is a small regularized parameter chosen to ensure the friction forces change continuously  with the body's velocity \cite{alben2019efficient}. This parameter is chosen from sensitivity analysis such that the results do not change more than $1\%$.

\subsection{Body actuation}
It is known that body actuation for snakes and limbless lizards that employ lateral undulation for locomotion is caused by muscle actuation that is propagated along its body from head to tail \cite{gasc1989propulsive}. This internal body actuation is modeled as a traveling wave of muscle torque given by,
\begin{equation}
\label{Eq: MuscleTorque}
\bm{Q}(s,t) = Q_\text{amp}\cos\left[\frac{2\pi\, n_w}{L}(s-c\,t)\right] \bm{b}
\end{equation}
where $L$ is the body length, $n_w$ is number of waves along the  body and $c$ is the wave propagation speed. For this study, $n_w$ is set to unity based on the previous finding that a curvature wavelength close to unity results in an optimized planar crawling locomotion \cite{alben2013optimizing}

\begin{figure}[h]
\centering
\includegraphics[width=1\columnwidth]{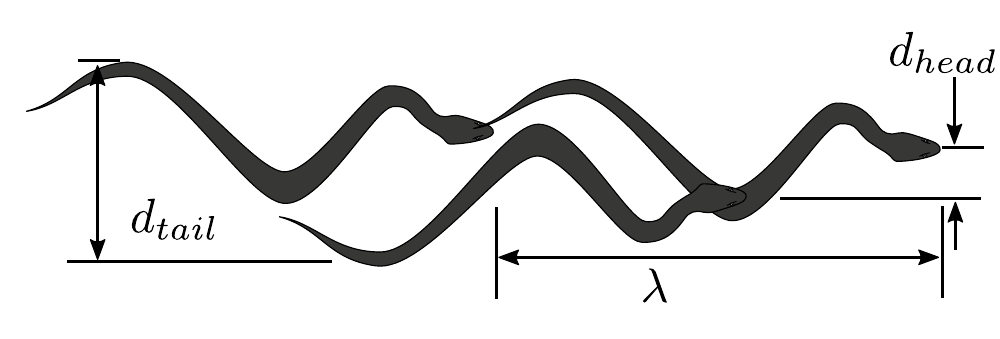}
\caption{For this work, shape ratio is the ratio of the tail-to-head lateral excursion}
\label{Fig: Shape Ratio}
\end{figure}

\subsection{Problem Setup}
\label{non-dim-section}
Equations \ref{a1-eq7} and \ref{eq:EnvFriction} are scaled by the body length $L$, body mass per unit length $m$, and the wave propagation speed of the muscle torque. 
The following nondimensional parameters quantify the effects of viscoelastic damping, body stiffness, and muscle torque amplitude,  
\begin{equation}
\label{piGroups}
\tilde \alpha = \frac{\alpha \,c}{L}, \hspace{0.5cm}
\tilde {EI}   = \frac{EI}{m\,L^2\,c^2}, \hspace{0.5cm}
\tilde Q     = \frac{Q_\text{amp} }{m\,c^2} 
\end{equation}
and the frictional environmental force is quantified with two nondimensional parameters below.
\begin{align}
\kappa_1=\frac{\mu_{\tau}}{\mu_n}, \hspace{0.5cm}
\kappa_2=\frac{g\,L\,\mu_n}{c^2} \label{non-dim_normal-friction}
\end{align}

As a nominal value for our study, we select a material stiffness $\tilde{EI} = 13$ and Torque amplitude $\tilde{Q} = 66$. These numbers are chosen to produce locomotion (for low damping case) with a traveling waveform with an amplitude to wavelength ratio of $\approx 0.2$ consistent with organisms that ``swim" in granular media as well as efficient organisms at low Reynolds number \cite{maladen2009undulatory,lauga2020fluid}. We note that the produced body curvature of this ``swimming" is $\approx 2$, which is lower than what is observed for terrestrial snakes whose body curvature could reach a maximum of $\approx 7$ \cite{hu2012slithering}. We explore the role of $\kappa_1$ while fixing the value of $\kappa_2= 7.8$ consistent with the friction values of snake scales traveling at normal speeds based on measured values from previous studies \cite{guo2008limbless,hu2009mechanics}.

It is well-known that various taxa, including terrestrial snakes that utilize undulation for locomotion, propagate waves of curvature from head to tail down their bodies. As in anguilliform locomotion, these bodywaves often increase  in amplitude from the head to the tail. We quantify this increased amplitude with a dimensionless ratio $d_\text{tail}/d_\text{head}$ shown in  figure \ref{Fig: Shape Ratio}. Here $d_{head}$ and $d_{tail}$ are the difference between the maximum and minimum lateral excursions of the head and tail, respectively.

At any given moment in time, a part of the friction density contributes to thrust, and the other part contributes to drag. To separate these, we defined two operators of $\mathcal{H}^+$ and $\mathcal{H}^-$ as,
\begin{align}
\mathcal{H}^+ (f) = \frac{f +|f|}{2},\hspace{0.5cm}
\mathcal{H}^- (f) = \frac{f -|f|}{2}
\end{align}
and define thrust and drag forces  acting on the swimmer along the locomotion direction  as,

\begin{align}
F_\text{thrust}(t) &=\int_0^L\mathcal{H}^+ ({\bm{F}_e}(s,t) \cdot \mathbf{t}_{\bar{\bm{v}}}) ds\label{eq:thrust}\\
F_\text{drag}(t) &=\int_0^L\mathcal{H}^- ({\bm{F}_e}(s,t) \cdot \bm{t}_{\bar{\bm{v}}}) ds\label{eq:drag}
\end{align}
where $\bm{t}_{\bar{\bm{v}}}$ it the mean travel direction. 

We quantify the efficiency of the crawler by comparing the work done by the thrust force to the total work done by the muscles. We address the time variance of the work and thrust by identifying the locomotion period and using the phase average of thrust work and total work. We call the distance traveled in one period of body shape variables a stride length $\lambda$.
Further, having established instantaneous thrust and drag functions, we can compute power and efficiency using the following equations:

\begin{align*}
P_\text{muscle}(t) &= \int_{0}^{L} \mathcal{H}^+ \left(\bm{\omega}(s,t) \cdot \bm{Q}(s,t)\right)\, ds\\
\\
\eta& = \frac{\overline{F_\text{thrust}} |\bar{\bm{v}}|}{\overline{P_\text{muscle}}}
\end{align*}
It is assumed that the swimmer cannot harvest power from the environment. Hence, the contribution to the power expenditure through the length is always non-negative. 
\section{Results}
\begin{figure}[h]
\centering
\includegraphics[width=1\columnwidth]{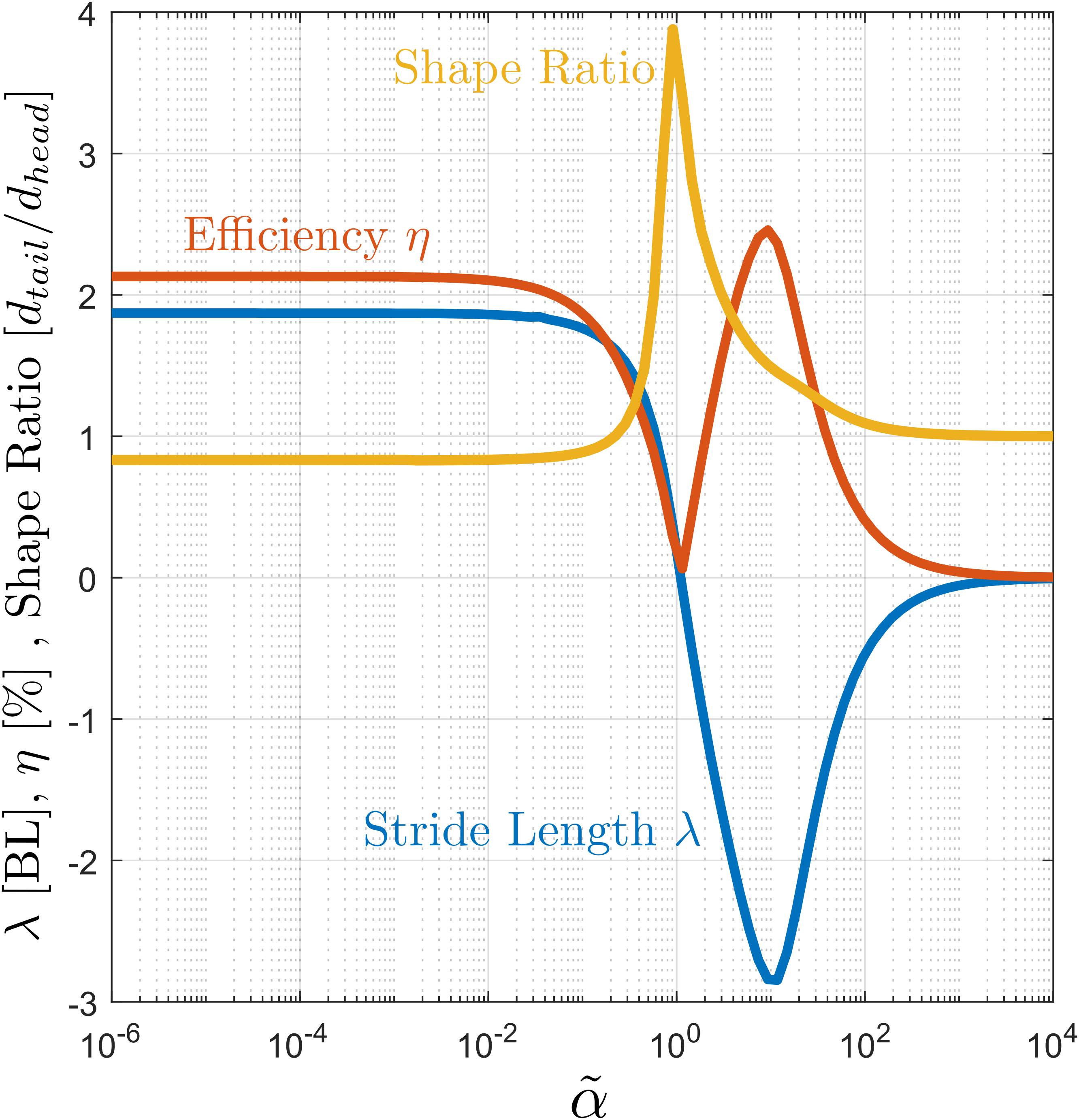}
\caption{Locomotion performance dependence on damping in a normal force dominated friction ($\kappa_1=\mu_{\tau}/\mu_n = 0$). Vertical axis is stride length $\lambda$, efficiency $\eta$ and shape ratio $d_{tail}/d_{head}$. }
\label{fig: slice}
\end{figure}

Figure \ref{fig: slice} shows computed steady locomotion results for different internal damping $\tilde{\alpha}$ values. Here the environmental friction is normal direction dominant with $\kappa_1 = 0$. We find that for low internal damping, each cycle produces motion opposite the muscle moment wave propagation direction. At the value of $\tilde{\alpha} \approx 1$, the crawler undulates in place with no net locomotion. Damping values beyond unity produce first increasing stride length in the direction of torque propagation and then quickly drop off to zero. The crawler achieves the maximum efficiency at the critical value of $\tilde{\alpha} \approx 10$. 

\begin{figure}
\centering
\includegraphics[width=0.8\columnwidth]{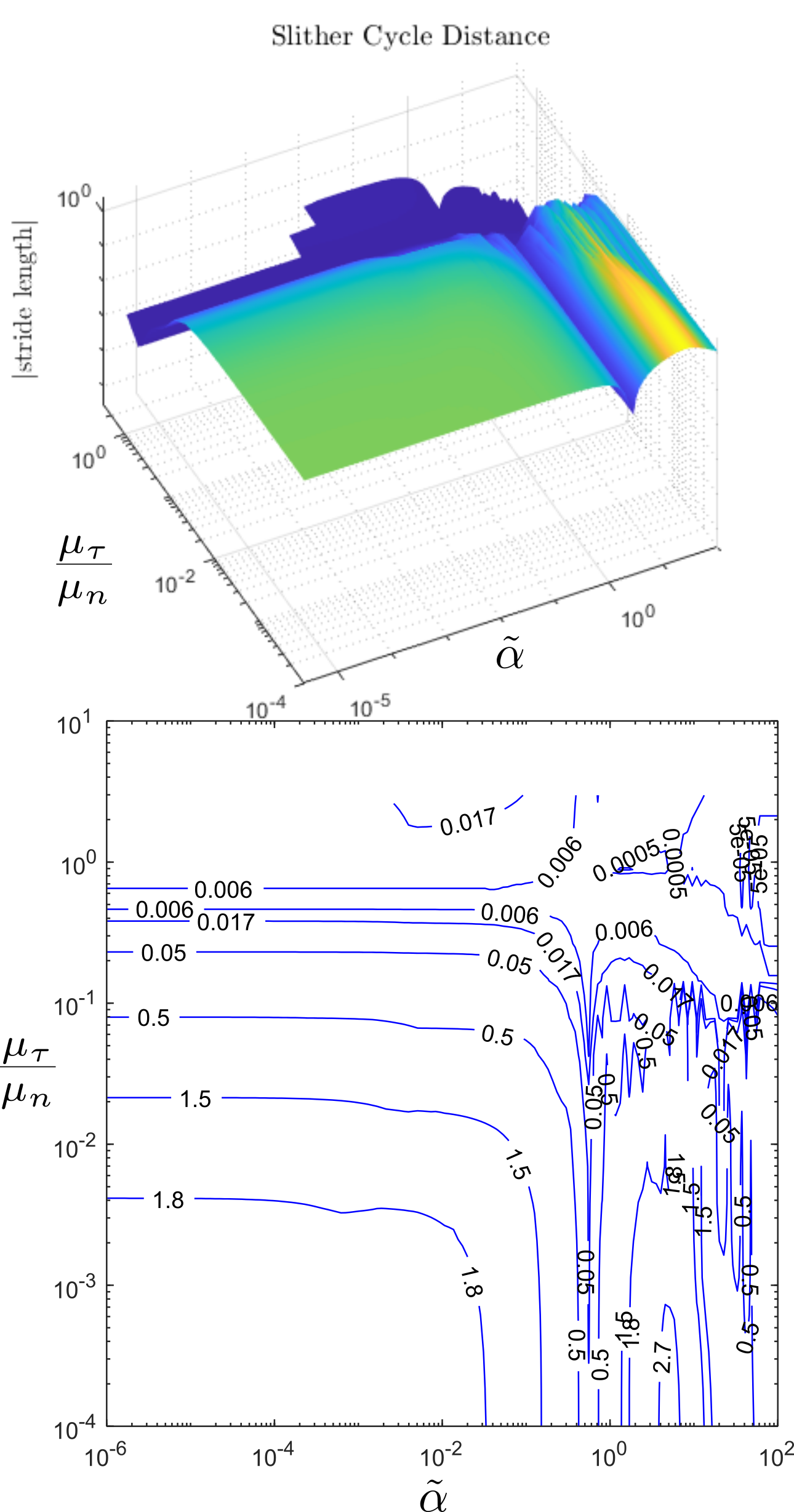}
\caption{Absolute value of stride length (distance traveled per cycle) as a function of damping and environmental friction 
}
\label{Fig: StrideFunctionFigure}
\end{figure}

\begin{figure}
\centering
\includegraphics[width=0.8\columnwidth]{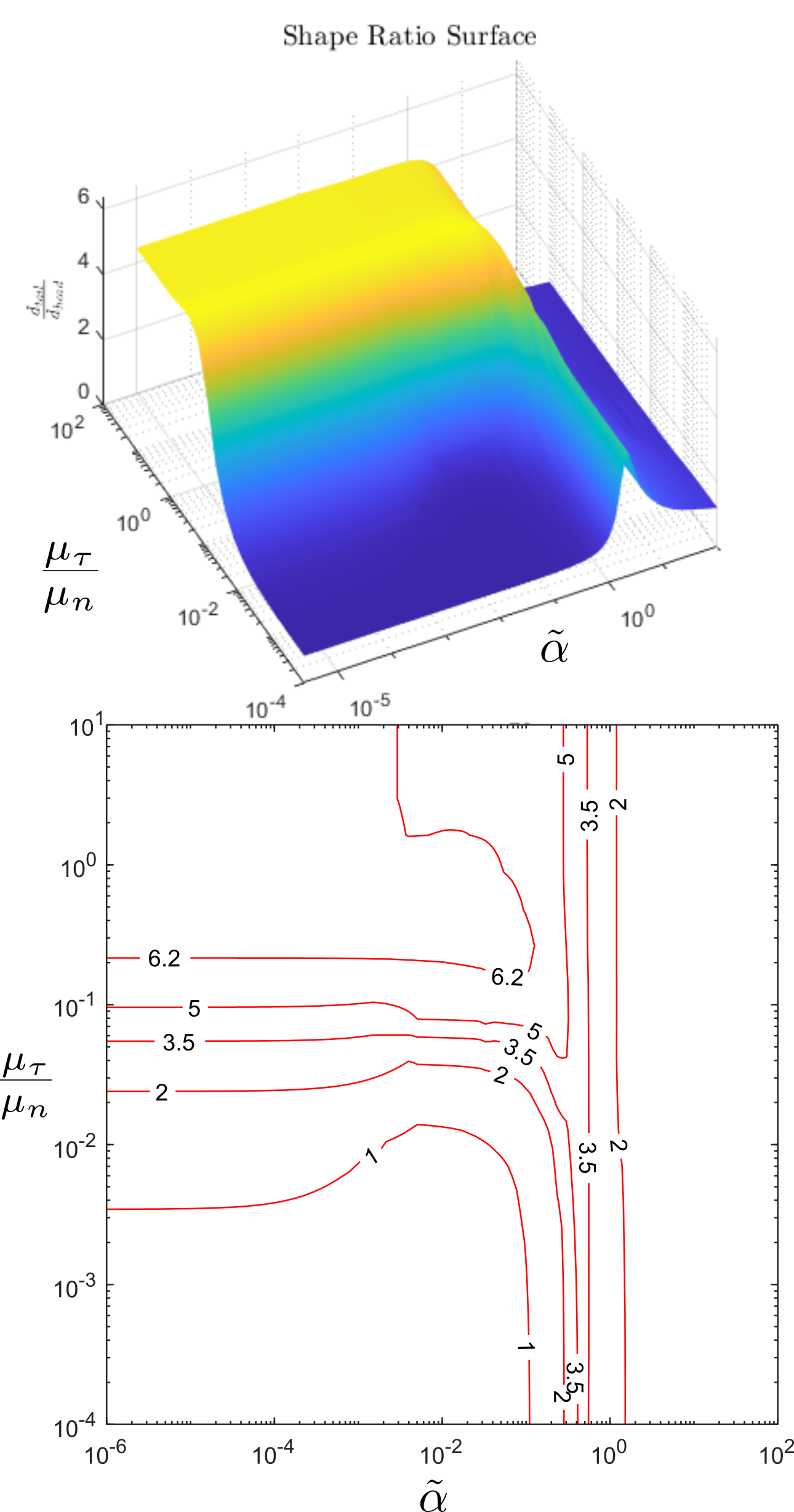}
\caption{Lateral Excursion ratio as a function of environmental friction and damping. We quantify that effect with the shape ratio or lateral excursion ratio ($d_{tail}/d_{head}$ as defined in figure \ref{Fig: Shape Ratio}).
}
\label{Fig: ShapeRatioFunction}
\end{figure}

In Figure \ref{fig: slice}, the shape ratio, $d_{tail}/d_{head}$, is around 0.8 for cases with low internal damping.  The Shape ratio increases to a maximum of $\approx 4$ near the critical damping condition  before returning to approximate unity as $\tilde{\alpha}$ increases. In other words, for sufficiently low or high damping cases, the head deviates approximately as much from the direction of travel as the tail. However, at the critical damping point, the tail deviates approximately four times as much as the tail. 

Expanding our study of the environmental effects and damping to a larger $40\, \times \,40$ computational grid, Figure \ref{Fig: StrideFunctionFigure} shows how the stride length modifies over a range of $ \mu_{\tau}/\mu_n$. On closer inspection, we find that there is a minimum in the parameter space at  \({\mu_{\tau}}/{\mu_n} \approx 0.5\), and a minimum for $\tilde{\alpha} \approx 1$. These local minima divide the parameter space into four quadrants. The global maximum of the stride length is obtained in the bottom right quadrant with higher damping and normal-dominant friction condition.  The second optimal stride length is obtained in the bottom left quadrant with low damping and  normal-dominant friction condition.  The top left quadrant has a much smaller local maximum for stride length that corresponds to high tangential friction and low damping. High damping and tangential-direction-dominant friction produce a global minimum in the top right quadrant.  

Figure \ref{Fig: ShapeRatioFunction} shows how the  shape ratio $d_{tail}/d_{head}$ changes with respect to $\tilde{\alpha}$ and $\mu_{\tau}/\mu_n$. Taking fig. \ref{Fig: StrideFunctionFigure} and \ref{Fig: ShapeRatioFunction} together, we see that a high shape ratio is associated with regimes with small stride lengths and often accompanied by no net locomotion. Conversely, a low shape ratio over the whole parameter space corresponds to longer stride lengths, except for the unique case of high tangent friction and high damping.

\begin{figure}[t]
\centering
\includegraphics[width=.9\columnwidth]{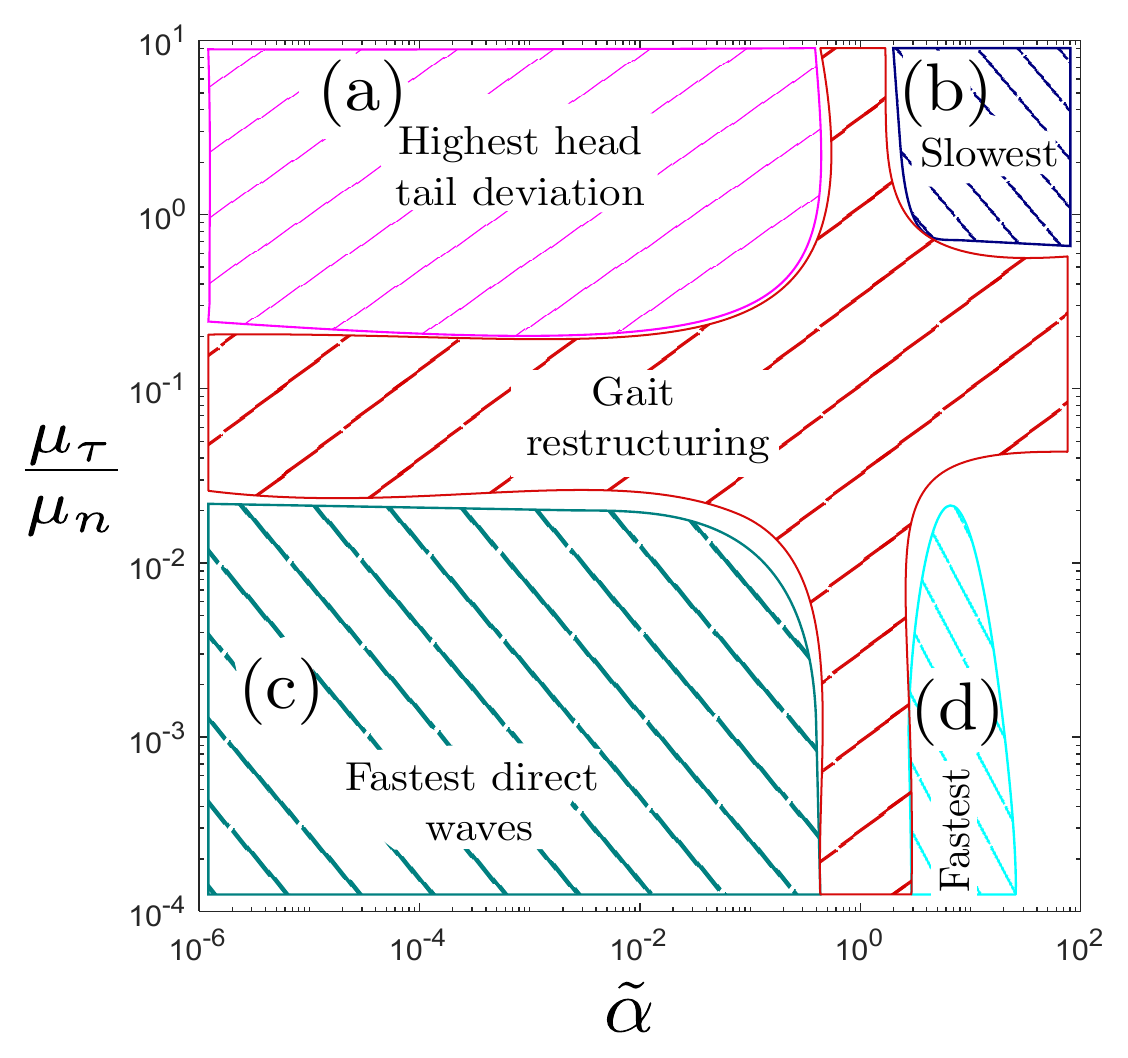}
\caption{Overview of behavior in various friction and damping regimes.}
\label{Fig: overview}
\end{figure}

Further summarizing the results, figure \ref{Fig: overview} labels areas of interest from Figures \ref{Fig: StrideFunctionFigure} and \ref{Fig: ShapeRatioFunction} in the friction, damping parametric space and Figure \ref{Fig: mode Shapes} presents the crawlers motion over one stride cycle in each region. Locomotion is slow in the tangentially dominated regions shown in fig. \ref{Fig: mode Shapes} (a) and (b); a salient difference is that the crawler body exhibits more lateral motion for the lower damping cases of (a). In  case (b), in contrast, the crawler undergoes approximate rigid body pitching rotation.

Figure \ref{Fig: mode Shapes} (c), and (d) illustrate the gait changes that cause the crawler to change direction as a function of damping. In subfigure (c), the crawler moves from left to right (red to blue), opposite the moment propagation direction. The different direction of motion is observed in subfigure (d), where it is shown that the crawler is moving from right to left in the same direction as the moment wave. A gait transition happens between left-to-right and right-to-left motion in the restructuring area. We note that in this region, the tail has a significant deviation outside the path the head follows. Figure \ref{Fig: mode Shapes} (d) offers the highest stride length gait. Here the amplitude of the crawler is significantly attenuated and characterized by a lower curvature amplitude and reduced bending wavelength.

The curvature ($\Omega_b$) contour plots in figure \ref{Fig: mode Shapes} show that in both the highly damped cases, i.e., (b) and (d), isoclines are increasing from left to right, showing that curvature waves are propagating from tail to head. However, we see the opposite isocline slope in (a) and (c).

The thrust distribution plots, ${\bm{F}_e}(s,t) \cdot \mathbf{t}_{\bar{\bm{v}}}$ in figure \ref{Fig: mode Shapes} show that region (a) is uniquely dysfunctional in that very high thrust is being generated but is being dissipated entirely by corresponding negative forces. Moreover, we can see that even though the stride of (d) is longer than (c), the locomotion generates lower forces. This is explained by the reduced gait waveform amplitude, which exposes the crawler to less drag force.

\subsection{Sensitivity Analysis}
\begin{figure}[b]
\centering
\includegraphics[width=0.9\columnwidth]{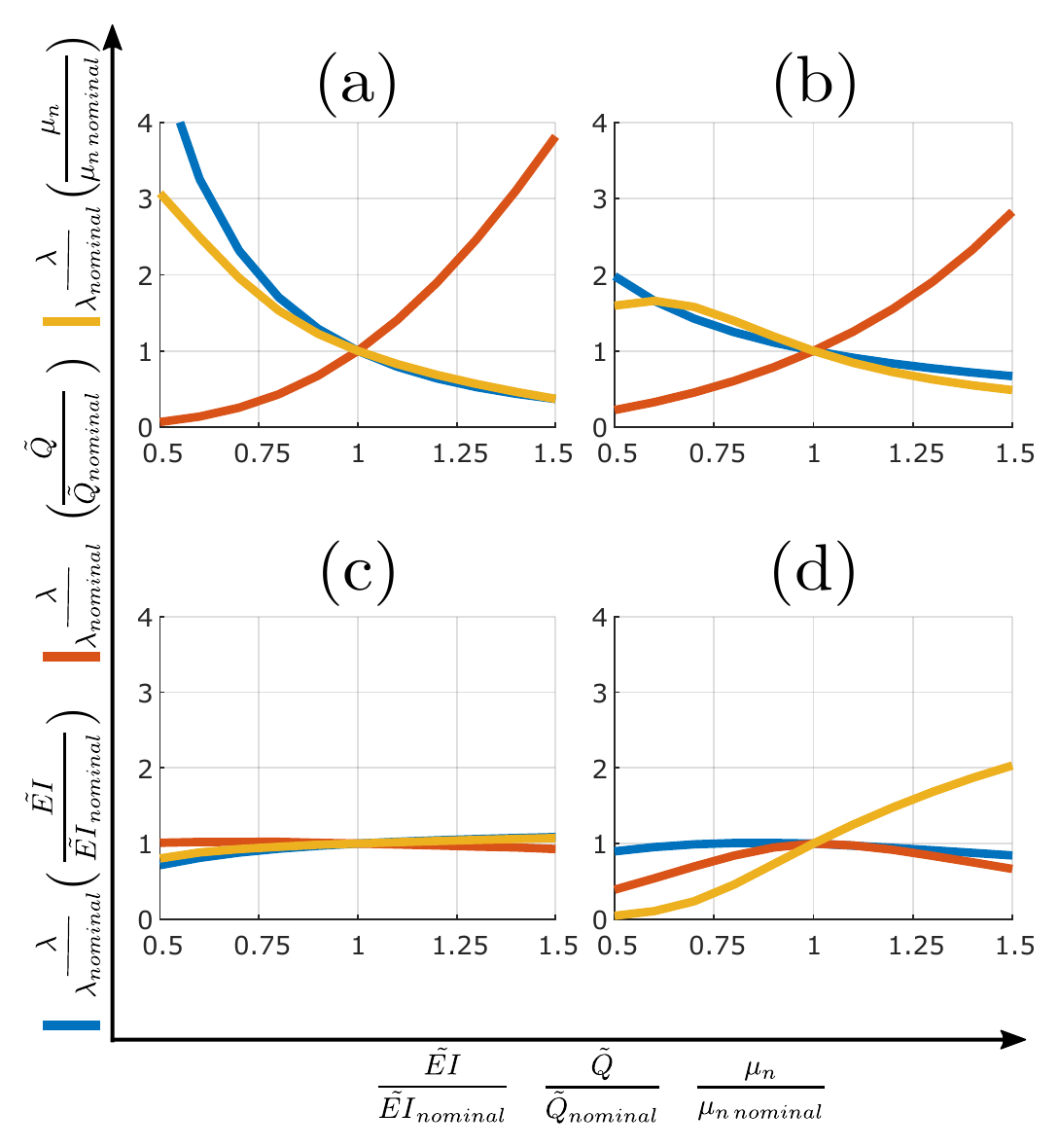}
\caption{
The sensitivity of different crawling gates in Figure \ref{Fig: overview}. The sensitivity of the stride length of a crawler to variation of friction is shown with yellow, the internal distributed torque with blue, and the body stiffness with red. }
\label{Fig: sensitivity}
\end{figure}

\begin{figure*}
\centering
\includegraphics[width=1.85\columnwidth]{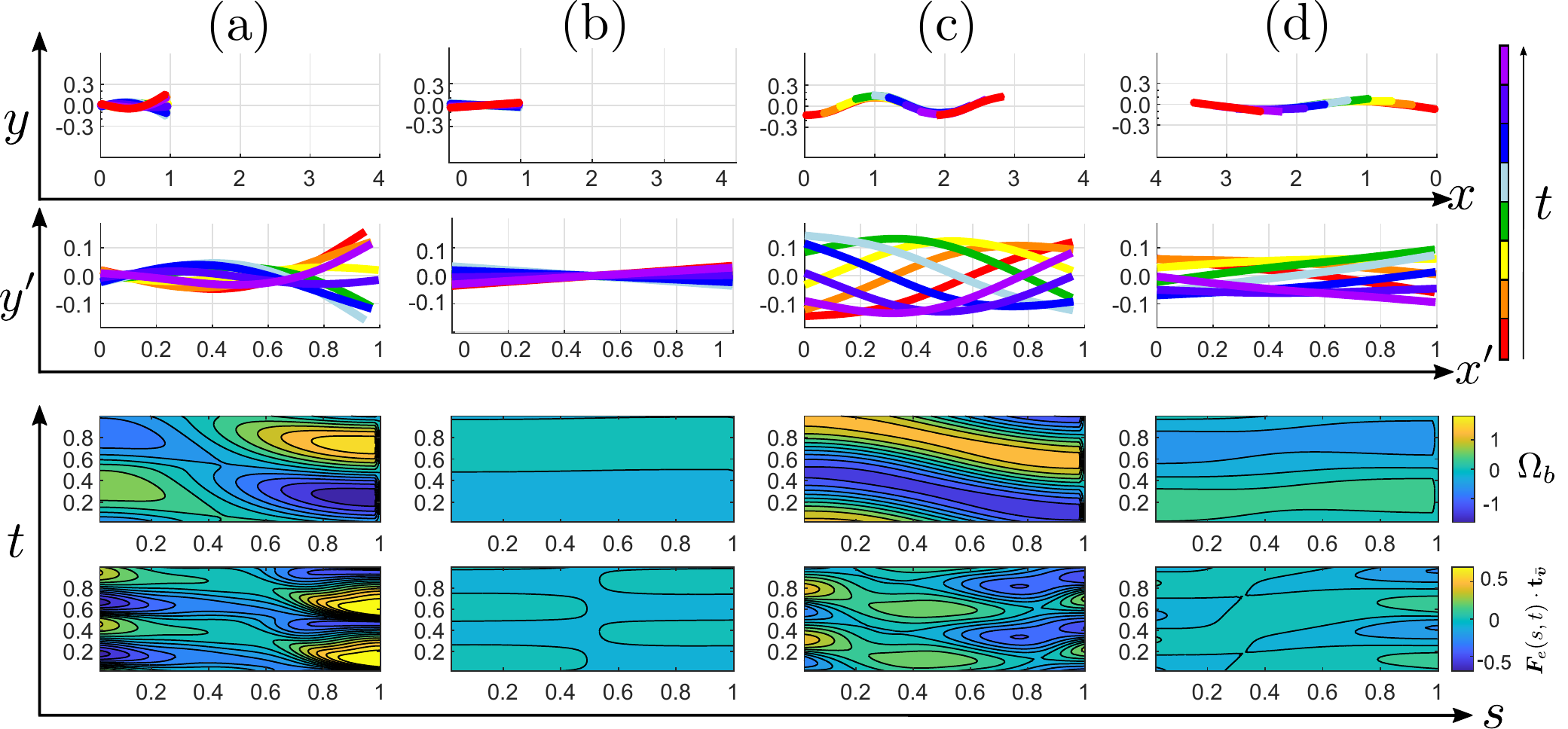}
\caption{Crawler trajectories in various regions of interest with labels (a) (b) (c) and (d) corresponding to Figure \ref{Fig: overview}. The first row shows crawler trajectories in a global$(x, y)$ reference frame. The second row shows the slithering body shapes for the regions of interest in a reference frame $(x',y')$ that moves with the body's average velocity. The third row shows how the curvature $\Omega_b$ develops along the body in space and time. The last row shows how the force distribution ${\bm{F}_e}(s,t) \cdot \mathbf{t}_{\bar{\bm{v}}}$ develops along the body in space and time. The high tangential friction of (a) and (c) is not conducive to locomotion by lateral undulation. Subfigures (c) and (d) show how damping changes the direction of locomotion for transverse-dominated friction regimes, such as wheeled snake robots.}
\label{Fig: mode Shapes}
\end{figure*}

The previous section showed the dependence of a crawler's gait on two parameters identified to be most influential for locomotion, the damping parameter $\tilde{\alpha}$ and the environmental friction ratio $\mu_{\tau}/\mu_n$. Figure \ref{Fig: sensitivity} shows the dependence of stride length on the remaining dimensionless groups: $\mu_n$, $\tilde{Q}$, and $\tilde{EI}$. This sensitivity analysis reveals that the fastest forward gait shown in fig. \ref{Fig: mode Shapes}(c) is not sensitive to variation in environmental friction, internal stiffness, or internal actuation. Likewise, the reverse gait shown in fig. \ref{Fig: mode Shapes}(d) is not sensitive to either internal torque or material stiffness. However, this fastest gait is sensitive to the normal friction coefficient, such that low normal friction can cause the crawler to come to a complete stop in this regime, and similarly, a 50\% increase can cause the stride length to double. For the transverse dominated friction regimes shown in \ref{Fig: overview} (a) and (c), an increase of bending moment or normal friction, or reduction of stiffness causes an increase in the stride length. This is the same trend that agrees with the effect of the $\mu_\tau/\mu_n$ ratio in figure \ref{Fig: StrideFunctionFigure}. 

\section{Discussion}

This work using the visco-elastic beam equations shows that lateral undulation is effective when normal or lateral friction is very high relative to tangential friction.  This finding is  consistent with the prevalence of wheels on snake robots that move using mainly lateral undulation since wheels provide a low value for $\mu_{\tau}/\mu_n$. This is also consistent with the findings of a numerical study by Hu et al. \cite{hu2012slithering} and the observation of Bruce Jayne that terrestrial snakes in isotropic friction environments tend to favor other gaits such as concertina over lateral undulation \cite{jayne1986kinematics}. 

Comparing the effect of damping in figure \ref{Fig: mode Shapes}, especially subfigures (c) and (d), we see that internal damping has the effect of stretching out the body of an undulating crawler. From a locomotion standpoint, this has the benefit of increasing stride length. It also changes the crawler's shape waveform by changing the crawling direction from opposite the torque wave propagation to moving in the same direction as the torque wave. This phenomenon could be used to control the direction of a robotic crawler by changing the internal damping during runtime, perhaps through a modulated aperture inside the crawler's body. Many mechanisms and fluidic circuitry can now be built into the material of a soft robot through the use of specialized manufacturing processes \cite{hubbard2021fully}. Here the external force is primarily in the normal direction to the surface, and the tangential friction is approaching zero. This friction regime is comparable to a wheeled snake robot.

Considering the motion on the bottom of the parameter space in figure \ref{Fig: StrideFunctionFigure} and \ref{Fig: ShapeRatioFunction}, we showed that when the forward resistance is sufficiently low, there is a benefit in using higher internal damping. The fact that locomotion is very slow in the tangentially dominated friction regions of figure \ref{Fig: mode Shapes} (a) and (b) is not surprising. Alben has demonstrated that a different curvature wave is optimal for locomotion in this friction regime. He also identified that efficient waves in this regime are still traveling waves but do not have a uniform amplitude along their spatial dimension see \cite{alben2013optimizing} figure 10 (compare Alben's $\mu_t=0.1$ and $\mu_t = 6$).

Figure \ref{Fig: mode Shapes} (c) and (d) shows that the low internal damping gait is such that the crawler's tail follows the trajectory of its head with a very slight deviation of the tail outside the path, and the whole body follows a traveling wave pattern. Thus, the shape ratio is almost a predictor of locomotion efficiency, such that shape ratios close to unity are often efficient crawlers. The sole exception is  region (b), where motion is practically entirely damped out.

When eel transition from swimming to crawling on land, the amplitude of the curvature of their gait increases. \cite{gillis1998environmental}. Gillis suggested that eels change their body actuation to produce these changes. Figure \ref{Fig: mode Shapes} (c) and (d) shows that damping reduces the amplitude of body curvature waves. This numerical result suggests that, similarly, the eel's body curvature amplitude reduction may be partly explained by the presence of damping (external in this case.) Thus, changes in muscle actuation may not be necessary to effect this curvature change.%

Considering these results in light of biological organisms, the stride length minima in the parameter space at  \(\mu_{\tau}/\mu_n \approx 0.5\) is roughly on the edge of the friction regimes measured for the scales of unconscious snakes (\(\mu_{\tau}/\mu_n \approx 0.6\)) \cite{hu2012slithering}. In this friction regime, we find that a traveling internal torque wave results in a traveling curvature wave of constant amplitude along the body, similar to figure \ref{Fig: mode Shapes} (c), but with a more lateral excursion. Alben et al. found an optimal gait in this comparable friction regime see figure 10 of \cite{alben2013optimizing}  with $\mu_{\tau} = 6$. Figure \ref{Fig: StrideFunctionFigure} shows that crawlers in this natural 
regime do not move faster with increased damping value and, thus, do not benefit from increasing internal damping. However, damping can still steer forward and backward motion in this regime. More work is needed here since the effects of scale angle activation and weight distribution are known to be significant for terrestrial snakes.

Normal or transverse-dominated friction is also comparable to organisms moving in sedimentary media. Some organisms on various length scales, such as razor clams and polychaete worms, can move forward and backward in sedimentary media\cite{che2010mechanics}. Recently biologically inspired robots have been developed that mimic this locomotion\cite{tao2020sbor}. Using internal damping as a control strategy to change the gait of a crawler in lateral force-dominated locomotion regimes might be useful in these types of soft robots to alter their burrowing direction. Considering the non-dimensional parameters of equation \ref{piGroups}, we can see that damping is a more significant factor at smaller length scales. Thus, this locomotion modification strategy is especially relevant for small robots.

\section{Conclusion}

The effect of viscoelastic properties on the locomotion properties of a flexible crawler is studied here. A traveling wave is employed as the body actuation, and the swimmer's response at different environmental friction conditions is quantified. The study reveals an optimal damping value for the crawlers with dominant normal friction forces. This condition can be leveraged to improve the performance of certain bio-inspired robots, such as wheeled snake robots, and systems that can adjust the normal friction coefficient through surface interface manipulation. The results reveal that the internal damping reduces the amplitude of a crawler's undulatory waveform and changes the direction of curvature propagation with an accompanied change in locomotion direction. The unique functionality of the internal damping in switching the direction of the locomotion can be used in future robotic design to enable a seamless transition between forward to backward motion by controlling internal damping.
\section*{Acknowledgments}
This work was funded by the National Science Foundation Expanding Frontiers in Research and Innovation Program, grant number 1935278.

\end{document}